\documentclass[review,number,sort&compress,3p,times,10.5pt,a4paper]{elsarticle}

\usepackage{lineno}
\usepackage{color}
\usepackage{multicol}
\usepackage{framed}
\usepackage[T1]{fontenc}  
\usepackage{textcomp}     
\usepackage{mathptmx}
\usepackage{url}
\usepackage{hyperref}
\usepackage{graphicx}
\usepackage{amssymb}
\usepackage{subfigure}
\usepackage{amsmath}
\newcommand{\angstrom}{\text{\normalfont\AA}}

\begin{document}
	
\begin{frontmatter}
\title{Structural and optical properties of CsI thin films: Influence of film thickness and humidity}

\author[1]{Nabeel Jammal}
\author[1]{Richa Rai}
\author[1,2]{Triloki}
\author[1]{B. K.~Singh\corref{mycorrespondingauthor}}
\cortext[mycorrespondingauthor]{Corresponding author}
\ead{bksingh@bhu.ac.in}

\address[1]{High Energy Physics Laboratory, Department of Physics, Institute of Science, Banaras Hindu University, Varanasi 221005, India}
\address[2]{Present address: Solid State and Structural Chemistry Unit, Indian Institute of Science, Bangalore 560012, India}

\begin{abstract}
			
Structural and optical studies have been performed on the thermally-evaporated "as-deposited" and "humid air aged" CsI thin films. The structural analysis for both "as-deposited" and "humid air aged" films shows a well-oriented peaks position of (110) and (220) lattice planes with a compressive stress in the films. The crystal quality has been investigated through the structural parameters. The increase in peak intensity as well as sharpness with film thickness implies the improvement of crystallinity. The optical absorbance of CsI films has been analyzed in the wavelength range of 190 nm - 900 nm in order to estimate the band gap energy of the films. Slater's model has also been used to explain the degradation of band gap energy with the increase in crystallite size.
			
\end{abstract}
		
\end{frontmatter}

\section{Introduction}
	
Cesium Iodide (CsI) thin films are widely used as photoelectron converters in particle physics and medical imaging experiments~\cite{Gratta,Mauro} due to its high quantum efficiency (QE) in the ultraviolet (UV) spectral range~\cite{Carruthers,Seguinot,Mine,Breskin}, a high secondary electron emission (SEE) yield~\cite{Schwarz,Gibrekhterman}, a relatively high stability in air among other alkali halides material~\cite{Sommer} and its capability to operate in a stable way with gaseous detectors~\cite{Dangendorf,Charpak}. In the recent past, a lot of research work has been done in order to enhance its QE~\cite{Nitti,B.K. Singh,J. Almeida} and several groups have made progress in this respect and a quantum efficiency of about 35-40\% at 150-160 nm has been reported for a 500 nm CsI thin film~\cite{Lu,we}. Under humid-free gas atmosphere, CsI thin film has been coupled successfully with a multiwire proportional chamber (MWPC) and various micropattern gas detectors such as gas electron multiplier (GEM), thick gas electron multiplier (THGEM)~\cite{Seguinot,Breskin2}, etc. For example, experiments such as COMPASS~\cite{P. Abbon} and ALICE~\cite{ALICE} at CERN both have implemented a reflective CsI thin films photocathode with a MWPC. However, PHENIX at RHIC~\cite{W. Anderson} has used a cascade of GEMs with a reflective CsI thin films for particle identification. Recently, dedicated R$\&$D efforts have been made to develop a cryogenic gaseous photomultipilers in which a triple layers of THGEM structure with a semitransparent layer of CsI film have been tested successfully under liquid xenon atmosphere for its use in dark matter search experiment~\cite{L. Arazi}. The detectors used for these experiments generally work in two modes; reflective (opaque) and semitransparent mode. The Quantum Yield (QY) of reflective films has been observed to be higher in comparison with the semitransparent films. This is because the semitransparent films have a small escape length (L$\leq$ 10 nm)~\cite{Lu}. Also, it has been revealed that the reflective films have more chemical stability against the moisture compared to the semitransparent films~\cite{Breskin2}. 
		
Keeping these facts in mind, the main issue related to CsI photocathode is the aging which causes a degradation in the photoemission properties and thus limits their life. In order to develop and optimize large area photocathodes based detector, the study of the optical and structural properties of the "as-deposited" and "aged" CsI photocathodes is ultimately required. In literature, there are several studies on CsI ageing, but its mechanism does not have a clear understanding yet. A. S. Tremsin et al.~\cite{A. S. Tremsin} have used TEM technique to study the structural properties of polycrystalline CsI thin film of 100 nm thickness under the impact of humid air and UV irradiation. M. A. Nitti et al.~\cite{493_(2002)_16–24} have examined the bulk structure and the surface of freshly and humid air exposed 100 nm CsI film by XRD and XPS measurements. Recently, Triloki et al.~\cite{Triloki} have studied the structural properties of thermally-evaporated CsI thin films of different thicknesses by using XRD and TEM techniques. C. Lu and K.T. McDonald~\cite{Lu} have previously reported the optical properties of freshly CsI thin films up to 31 nm thickness. This work is an extension of the previous works, where we have systematically studied the structural and optical properties of semitransparent and reflective CsI thin films grown under similar vacuum-environment in case of "as-deposited" and "humid air aged".

\section{Experimental Details}

\subsection{Deposition of CsI thin films}
	
The deposition of reflective (500 nm and 300 nm) and semitransparent (50 nm and 30 nm) CsI thin films was done in a stainless steel 18" diameter spherical vacuum chamber. Prior to the deposition process, the substrates were cleaned by treating with distilled water, acetone, and alcohol to remove any undesirable impurities. CsI powder of 5N purity from ALFA AESAR was placed in a tantalum (Ta) boat. The substrate was kept at approximate 20 cm distance from the Ta boat. After that, a high vacuum (of the order of $10^{-7}$ Torr) inside the chamber was created by the means of a turbo molecular pump of pumping speed 510 L/S (Pfeiffer Vacuum, Germany). In order to evaporate the CsI powders, a high current ($\sim$80 Amp) was applied by using a step-down transformer and a current controller (Morefield UK). The deposition rate and film thickness were monitored by a quartz crystal-based thickness monitor (Sycon STM 100). The deposition rate was kept at 1-2 nm/s.  After deposition, the prepared films were extracted to a small desiccator under constant flow of nitrogen ($N_2$) gas to keep it away from interactions with the humidity. The desiccator was pumped for few minutes by a small diaphragm pump. Then immediately, the prepared films were transported to the characterization setup. 

\subsection{Characterization}
	
The structural parameters of CsI thin films deposited on stainless steel (S.S) substrate were studied by X-ray diffraction (XRD) technique in the Bragg-Brentano para focusing configuration ($\theta/2\theta$ geometry) using PANalytical X'Pert PRO XRD system. The incident beam optics consists of a CuK$\alpha$ radiation source ($\lambda$ = 1.5406 \AA). XRD measurements were performed in continuous scan mode in the range $2 \theta = 10^\circ - 80^\circ$ at room temperature. The diffracted beam optics consists of a 0.04 rad solar slit and a scintillator detector. A 40 kV voltage and 30 mA filament current were applied on the emitted electrons from the cathode filament to accelerate it towards the anode plate (Cu). For each characterization, two samples were prepared; one of them was scanned just after deposition and the other exposed to the humid air ($RH = 60\pm5\%$) for one hour before scanning.
	
The optical properties of CsI thin films were carried out by using Ultraviolet-Visible (UV-Vis) spectrometer from a Perkin Elmer UV-Vis spectrometer (model: $\lambda$ 25). In this UV-Vis spectrometer, two kinds of lamps were used; a deuterium lamp in order to cover the ultraviolet (UV) range and a halogen lamp in order to cover the visible (Vis) and the near infrared (NIR) range. The Perkin Elmer ($\lambda$~25) spectrometer was used to measure the absorbance of the samples in the wavelength range of 190 nm - 900 nm with a scan step size of 1 nm and the wavelength accuracy is $\pm$ 0.1 nm~\cite{Manual}. In this study, all the films were deposited on quartz (Qz) substrate due to its transparency in this wavelength range. The optical absorbance of CsI films in case of "as-deposited" and "humid air aged" ($RH = 60\pm5\%$) for 1 hour, 2 hours and 24 hours was performed. Due to almost similar results in the optical properties for short time exposure (up to 2 hours), we only present here the result after 24 hours of humid air exposure.

\section{Results and discussion}
	
\subsection{\textbf{Structural properties of CsI thin films}}

The XRD patterns of "as-deposited" and "humid air aged" CsI films are shown in Fig. 1. The sharpness in XRD peaks of CsI films indicates its crystalline nature. The XRD patterns of 500 nm thick CsI film exhibit the most intense peak at (110) lattice plane, followed by several peaks at (211), (220), (222) and (321) lattice planes can be clearly observed in both cases ("as-deposited" and "humid air aged"). For 300 nm thick CsI film, we may observe the peaks at (110), (220) and (321) lattice planes in both cases. For 50 nm thick CsI film, we may observe only two peaks at (110) and (220) lattice planes in both cases. For 30 nm thick CsI film, we may observe just the peaks at (110) and (220) lattice planes in case of "as-deposited" CsI film, and one more peak at (310) lattice plane can be seen in case of "humid air aged" CsI film. The lattice plane assigned to the most intense peak (110) for both cases corresponds approximately to the Bragg's angles $2\theta \approx 27.8^\circ$, however in the case of other lattice plane (220),  Bragg's angles is about $2\theta \approx 57.1^\circ$. The lattice planes corresponding to the CsI powder are: (110), (200), (211), (220), (310), (222), and (321). All the appeared XRD peaks are attributed to body-centered cubic (bcc) structure. Furthermore, we went for thicknesses less than 30 nm, but we did not observe any lattice plane. Based on the appeared XRD peaks shown in Fig. 1, it can be seen that (for both cases) the number of XRD peaks, sharpness in peaks and their intensity increase with the increase in film thickness. This increase in peak intensity as well as sharpness in peaks imply the improvement of crystalline quality of these films. Also it can be observed that all CsI films show almost similar increase of the peak intensity with the film thickness before and after exposing to humidity.

\begin{figure}[!h]
		\centering
		\includegraphics[width=7cm,height=10cm]{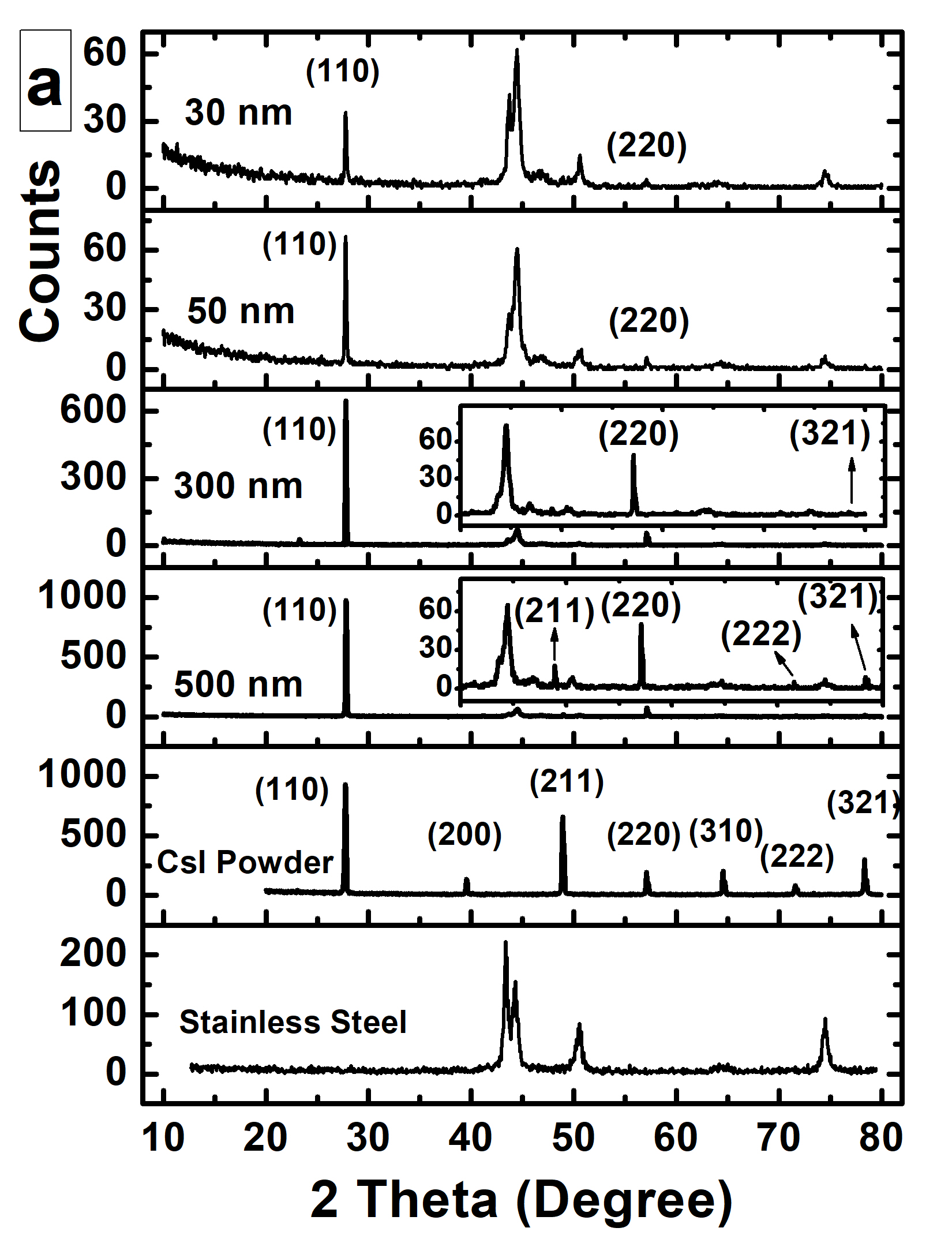}
		\hspace{0.5 cm}
		\includegraphics[width=7cm,height=10cm]{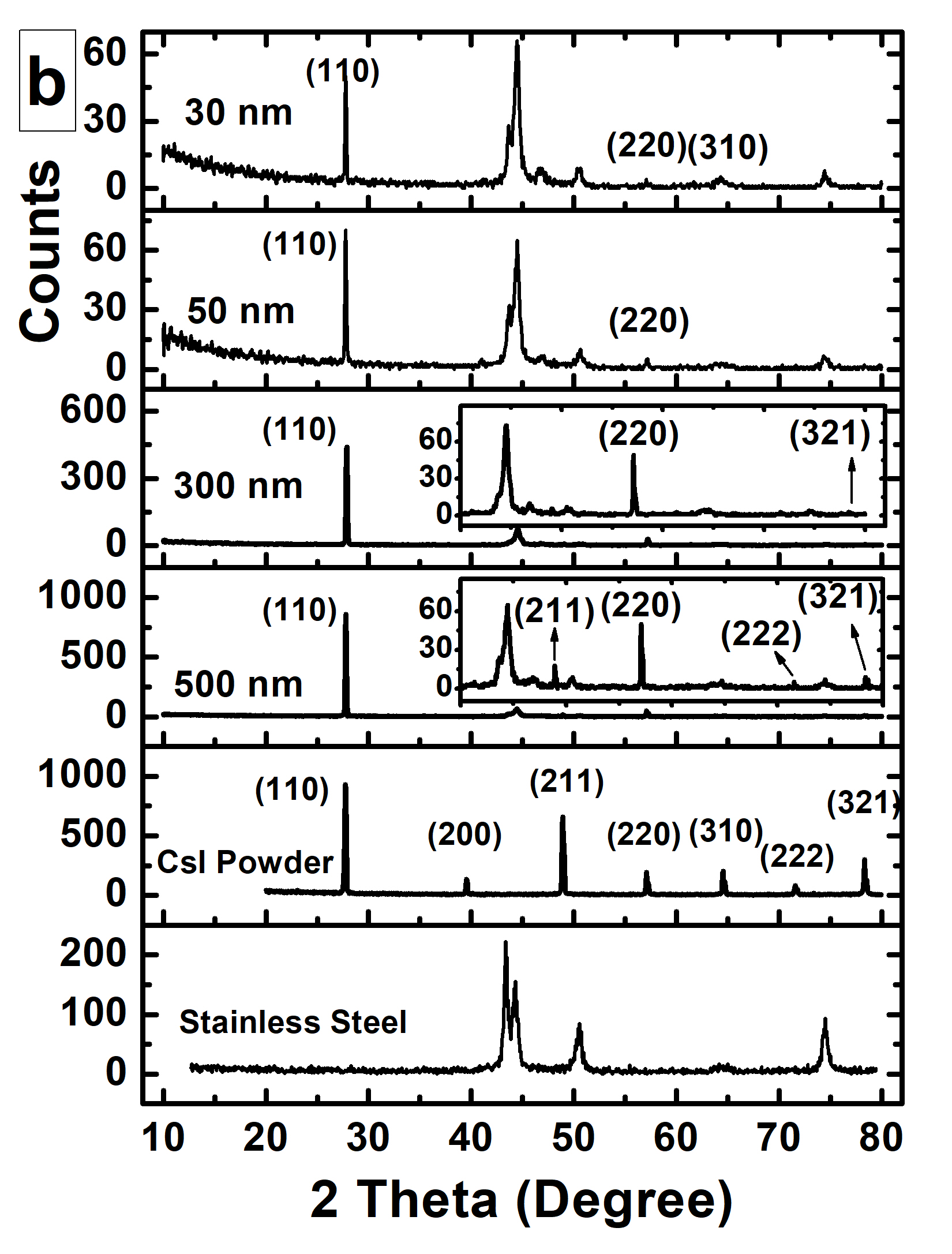}
		\caption{X-ray diffraction pattern of (a): "as-deposited" and (b): "humid air aged" CsI thin films of different thicknesses (reflective and semitransparent), deposited on S.S substrate.}
\end{figure}
	
Fig. 2a shows the full width at half maximum ($2\theta$-FWHM) of CsI thin films as a function of film thickness for both cases ("as-deposited" and "humid air aged"). It can be observed that the $2\theta$-FWHM of (220) lattice plane is higher than the $2\theta$-FWHM of (110) lattice plane in case of semitransparent films and almost same in case of reflective films. In the case of "as-deposited" CsI films, $2\theta$-FWHM decreases from $0.15^\circ$ to $0.10^\circ$ for (110) lattice plane and from $0.47^\circ$ to $0.07^\circ$ for (220) lattice plane when the film thickness increases from 30 nm to 500 nm. Also in case of "humid air aged" CsI films, $2\theta$-FWHM decreases from $0.14^\circ$ to $0.10^\circ$ for (110) lattice plane and from $0.94^\circ$ to $0.05^\circ$ for (220) lattice plane when the film thickness increases from 30 nm to 500 nm. The decrease of $2\theta$-FWHM may be understood in terms of the crystallite size and the strain developed in the film. Fig. 2b shows the correlation between $2\theta$-FWHM and crystallite size of the films. The crystallite size (D) of CsI thin films has been calculated by using the classical Scherrer's formula~\cite{10}. It can be noticed that the $2\theta$-FWHM decreases with the increase in crystallite size for both (110) and (220) lattice planes. It is believed that the main reason in the decrease of $2\theta$-FWHM with the increase in film thickness is possibly due to the increase in crystallite size of the films~\cite{Bin}, as shown in Fig. 2b. Thus, this reveals an improvement of the films crystallinity. 
	
	\begin{figure}[!ht]
		\centering
		\includegraphics[scale=0.28]{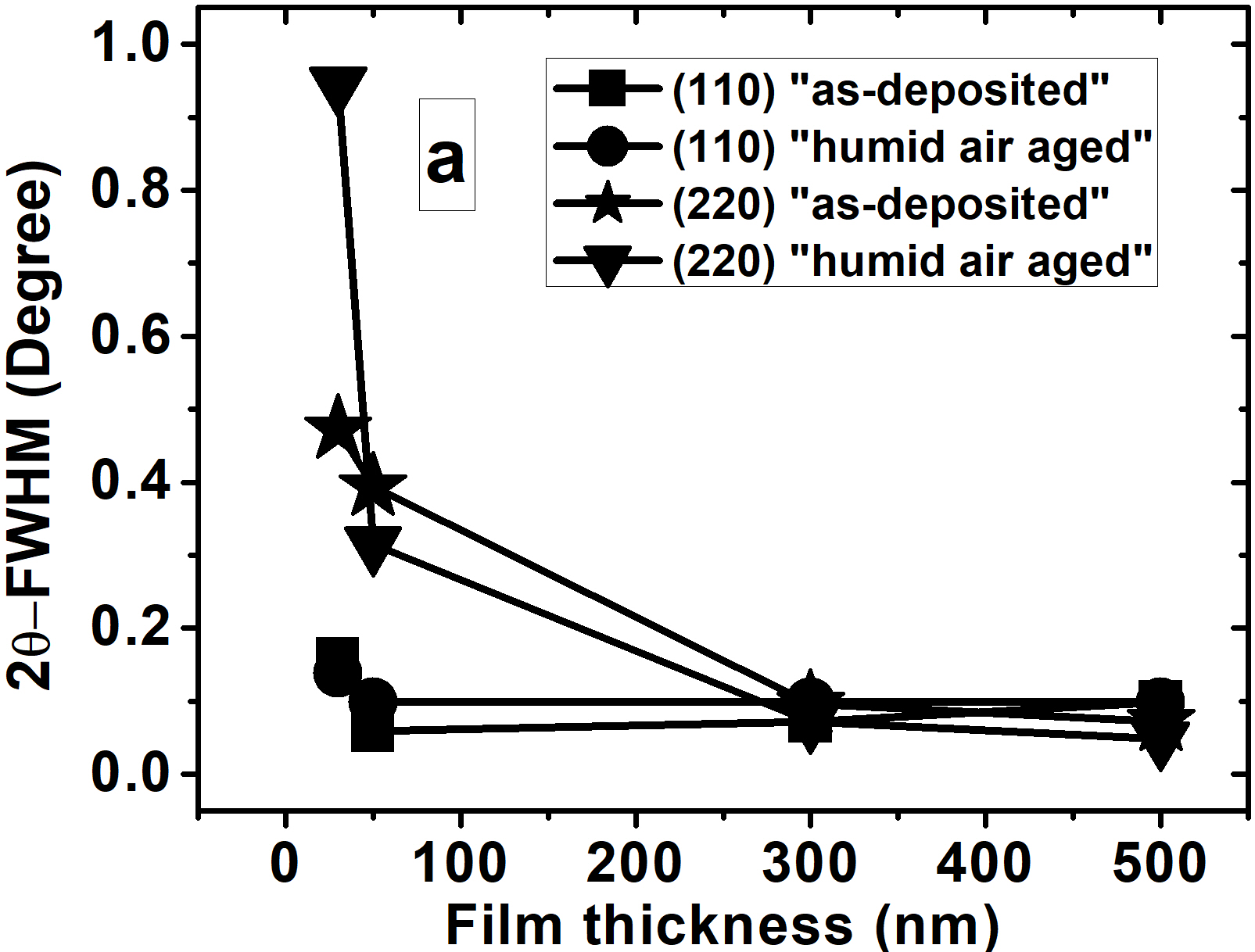}
		\hspace{0.5cm}
		\includegraphics[scale=0.28]{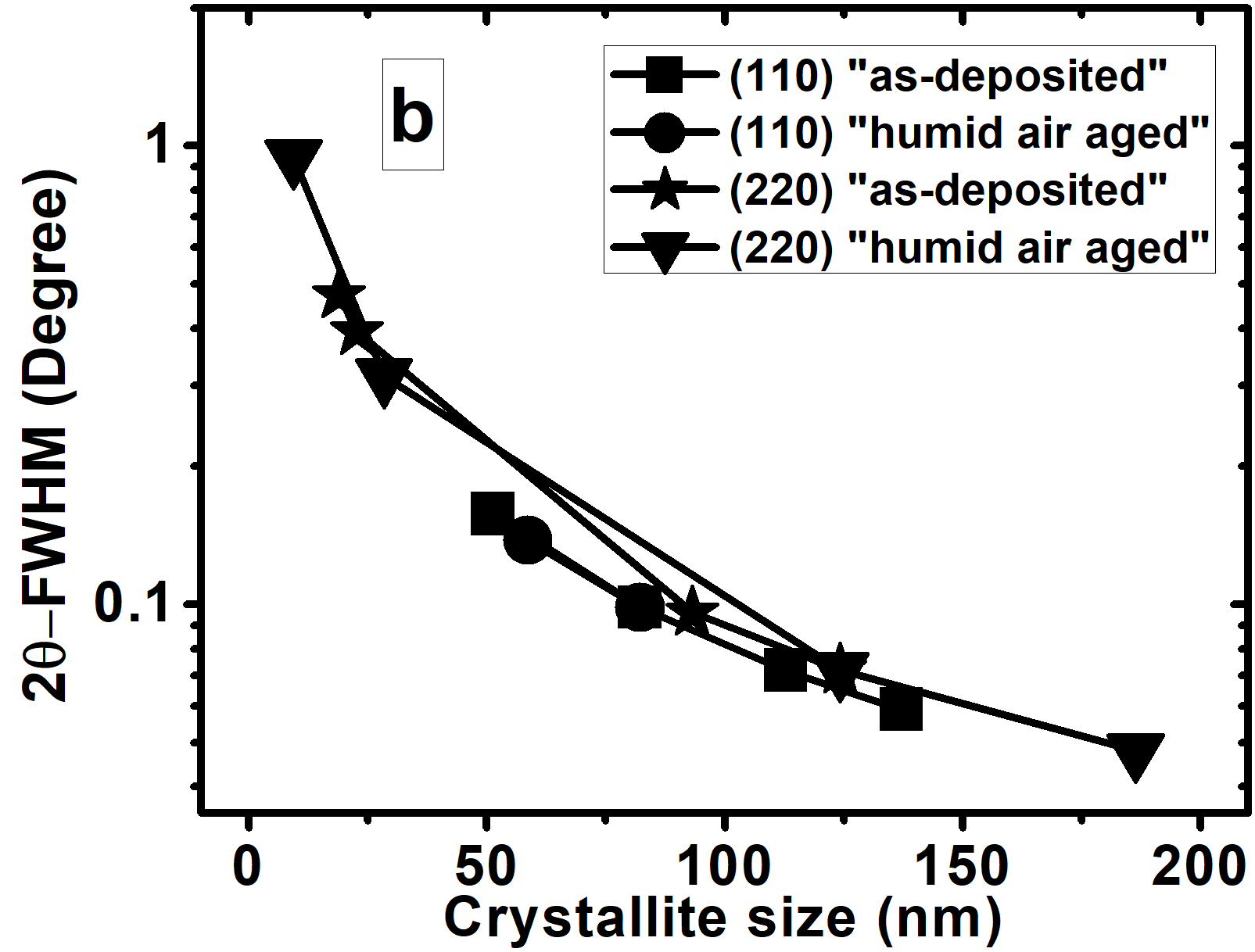}
		\caption{$2\theta$-FWHM of (110) and (220) lattice planes of "as-deposited" and "humid air aged" CsI thin films as a function of (a): film thickness and (b): crystallite size.}
	\end{figure}
	
The relationship between peaks position (2$\theta$) of (110) and (220) lattice planes with film thickness is illustrated in Fig. 3. In the case of "as-deposited" CsI films, the value of (2$\theta$) increases from $27.78^\circ$ to $27.80^\circ$ for (110) lattice plane and from $57.09^\circ$ to $57.14^\circ$ for (220) lattice plane with the increase in film thickness from 30 nm to 500 nm. Also in case of "humid air aged" CsI films,  the value of (2$\theta$) increases from $27.76^\circ$ to $27.77^\circ$ and from $57.00^\circ$ to $57.11^\circ$ for (110) and (220) lattice planes respectively with the increase in film thickness from 30 nm to 500 nm. In comparison with the peaks position of CsI powder ($2\theta~of~(110)=27.72^\circ,~2\theta~of~(220)=57.07^\circ$, shown with a sharp solid line), one can observe that CsI films peaks position have a small shift to the higher angles of $2 \theta$ (to the right). This shifted peak position corresponds to the interplanar spacing (d$_{(hkl)}$) of CsI films ($d_{(110)}\approx3.211~\angstrom$, $d_{(220)}\approx1.611~\angstrom$) which is smaller than the CsI powder ($d_{(110)}=3.124~\angstrom$, $d_{(220)}=1.612~\angstrom$), indicating a compressive stress acting in the film.  It is recognized that the shift to the higher angles of $2 \theta$ indicates a decrease in d-spacing which means a compressive stress acting in the film, and in contrast, the shift to the lower angles of $2 \theta$ indicates an increase in d-spacing which means a tensile stress acting in the film.

\begin{figure}[!ht]
		\centering
		\includegraphics[scale=0.15]{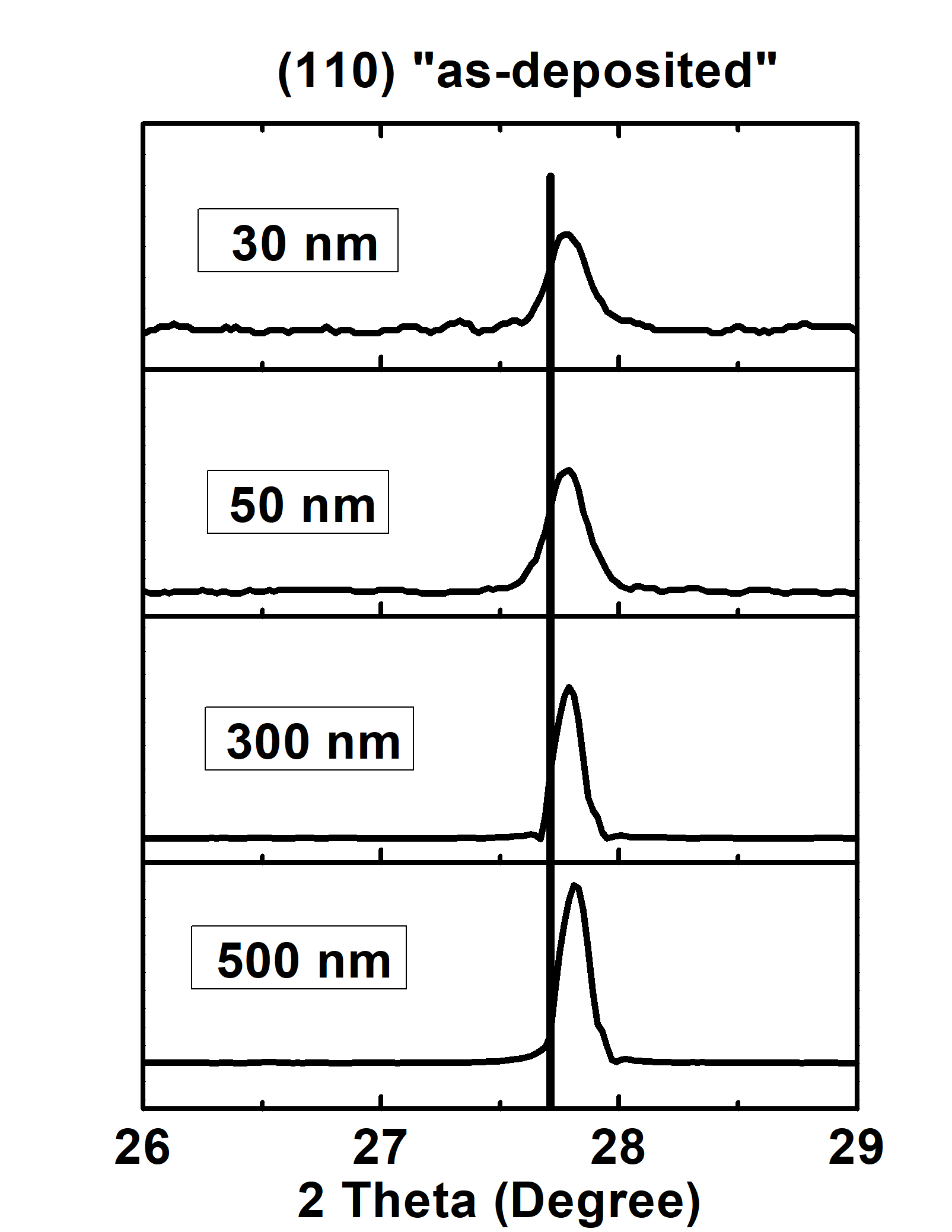}
		\hspace{0.5cm}
		\includegraphics[scale=0.15]{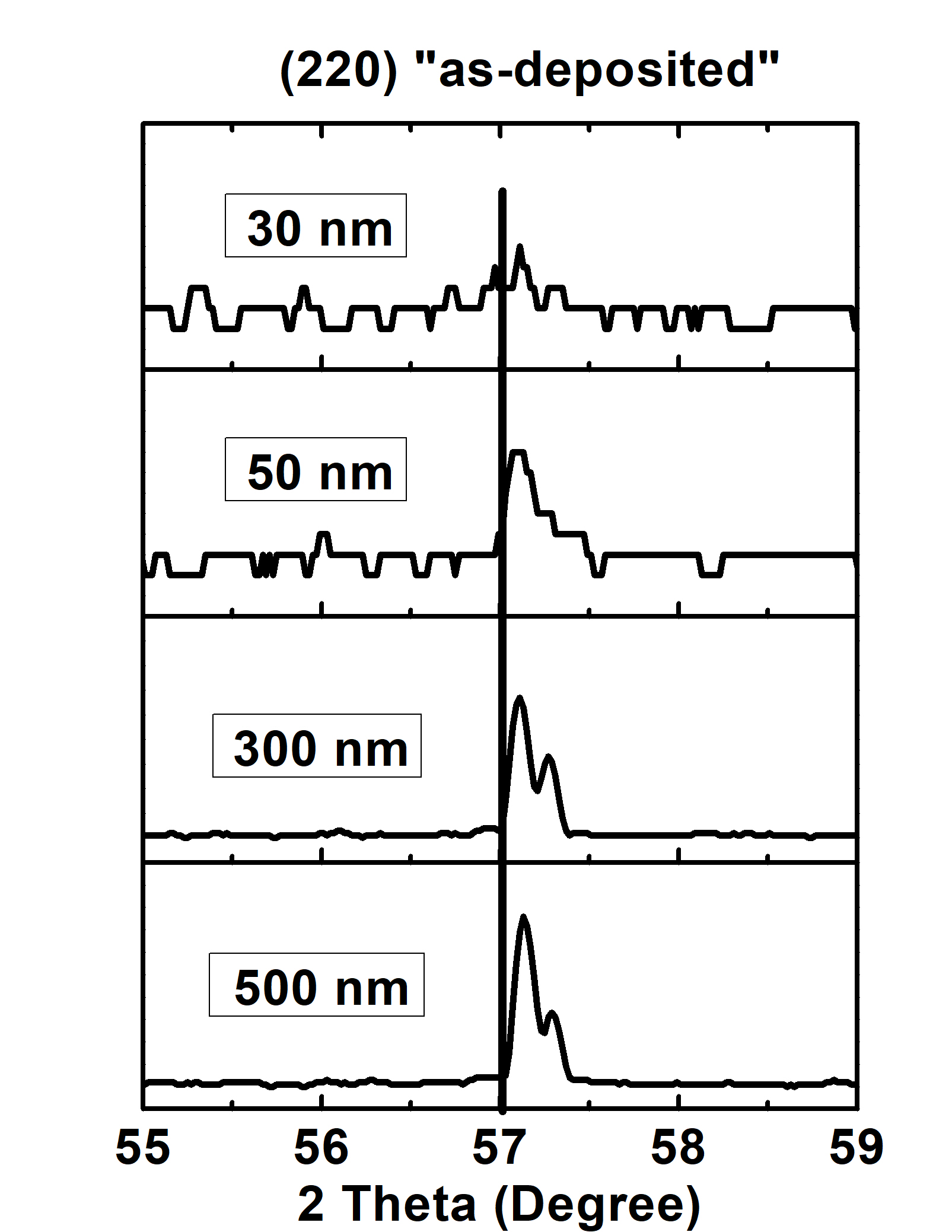}
		\vspace{0.5cm}
		
		\includegraphics[scale=0.15]{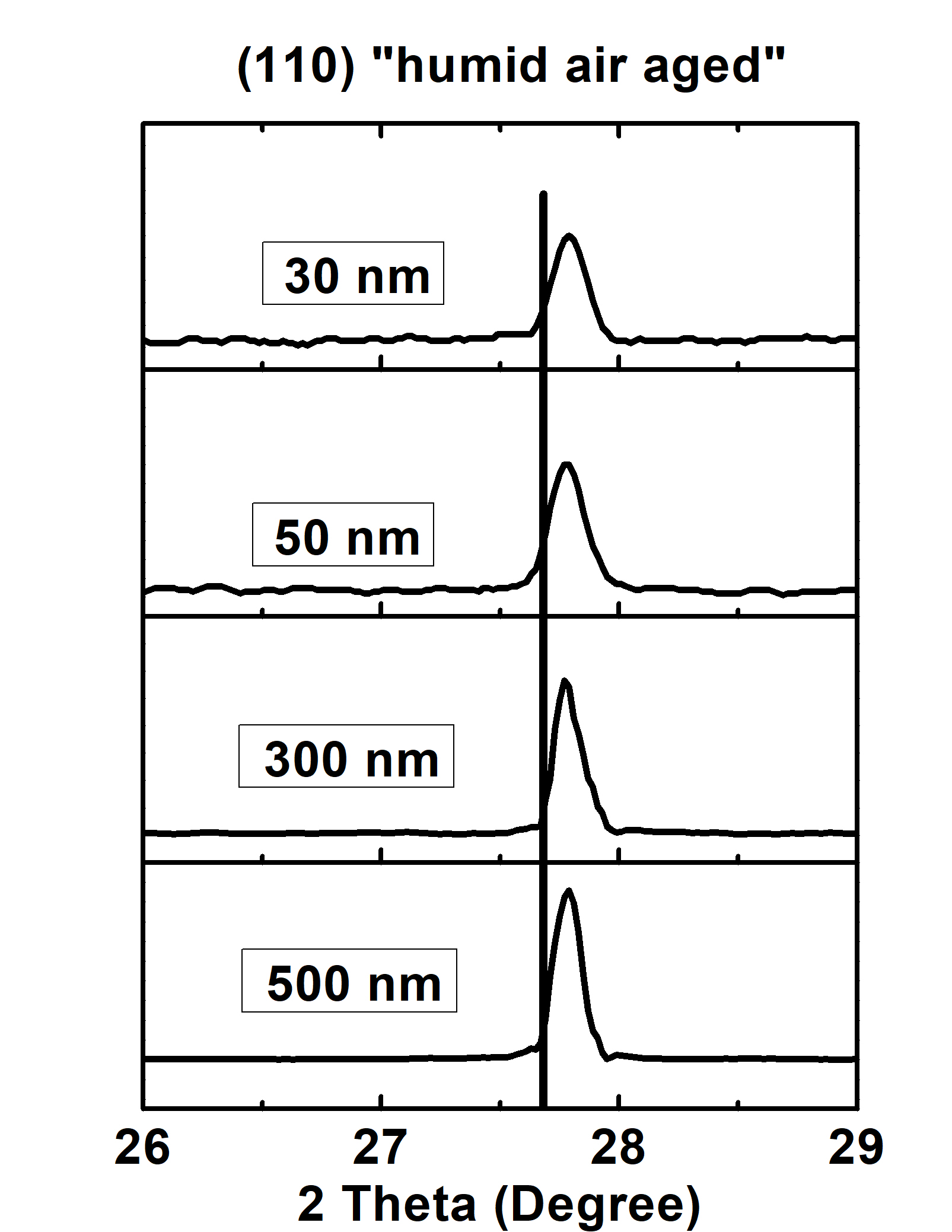}
		\hspace{0.5cm}
		\includegraphics[scale=0.15]{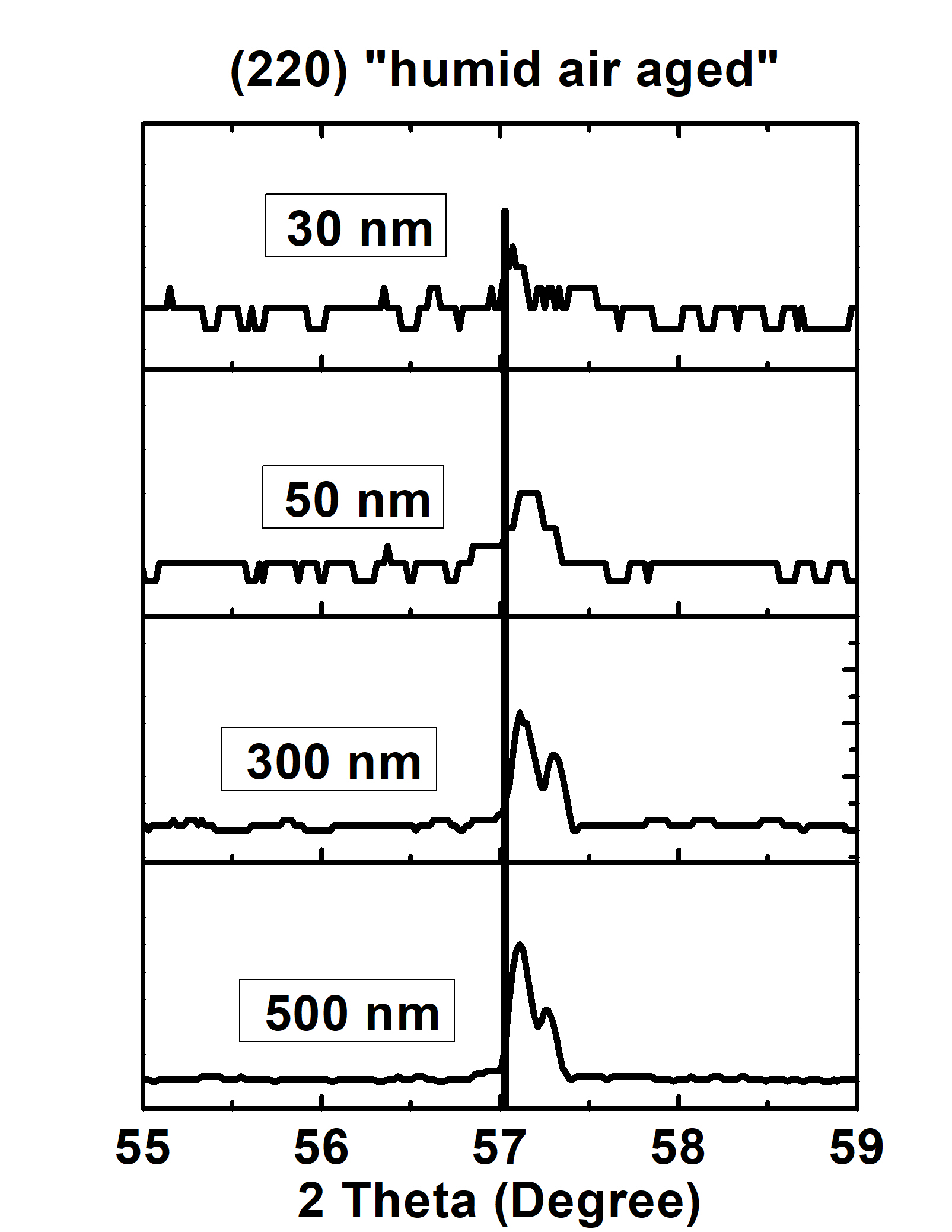}
		
		\caption{The shift in the (110) and (220) peaks of the X-ray diffraction pattern as compared to a CsI powder (shown with sharp solid line) before and after exposure to humidity.}
	\end{figure}
	
These stresses acting in the film may appear during the film preparation and can cause significant effects on the properties of the materials. Therefore, understanding the stress behavior in the CsI films is required. The inter-planner spacing of CsI powder ($d_{stand}$) and the experimentally obtained values of inter-planar spacing ($d_{exp}$) can be used to determine the strain in the films as it is illustrated in equation (1):

\begin{equation}
\label{eq:emc}
\mathbf{strain=\frac{d_{stand}-d_{exp}}{d_{stand}}}
\end{equation}
then the stress can be quantified by multiplying the strain $\Delta$d/d, where $\Delta d = d_{stand}-d_{exp}$, by the elastic constant of material. Fig. 4 shows the variation of strain developed in the CsI films as a function of film thickness. It can be observed that the compressive strain in both cases ("as-deposited" and "humid air aged") is slightly increased with the increase in film thickness. Also, the compressive strain in (110) lattice plane is observed to be stronger in comparison with the compressive strain in (220) lattice plane for both cases. In the case of low melting materials such as CsI having high material mobility, film stress is considerably small and both tensile and compressive stresses may build-up with increasing film thickness. On the basis of our experimental results on 30 nm - 500 nm "as-deposited" and "humid air aged" CsI thin films, we have observed a compressive stress throughout the films. We may believe that the compressive stress may originate from compressive strain from film-substrate interface during the film growth due to surface-tension effect~\cite{52_(1978)_215-229,171_(1989)_5-31}. As the film thickness is further increased, we may believe that this strain is communicated through the whole film and as a result, presence of compressive stress continue to increase with the film thickness. In a recent work of K. Kumar et al.~\cite{19}, they have observed tensile stress for "as-deposited" CsI films with thicknesses 580 nm - 730 nm. However they also observed a compressive stress below 580 nm  CsI films which is in agreement with our findings. It can be observed that the impact of humidity on the strain values is very slightly for all films, where it is in the same range of the "as-deposited" one. Therefore, we may say that the stress acting in the films is affected by some other factors. A. M. Engwall et al.~\cite{A. M. Engwall} have investigated the residual stress developed in the thin films and mentioned that in addition to the processes of film deposition, the stress can be also arisen by effects that may happen after deposition, such as thermal expansion mismatch between the film and substrate or grain growth in films that have sufficient material mobility. 

\begin{figure}[!ht]
	\centering
	\includegraphics[scale=0.28]{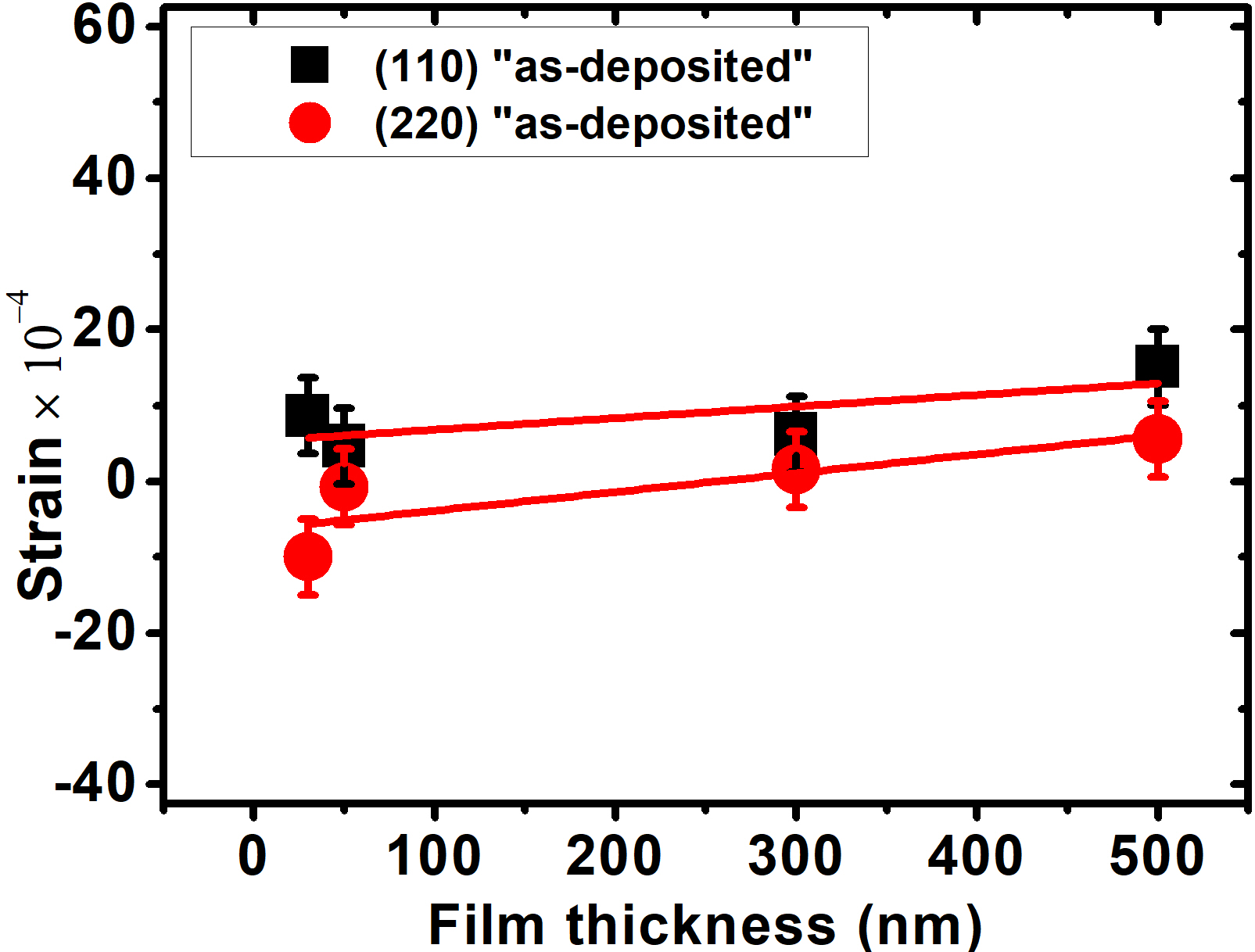}
	\hspace{0.5cm}
	\includegraphics[scale=0.28]{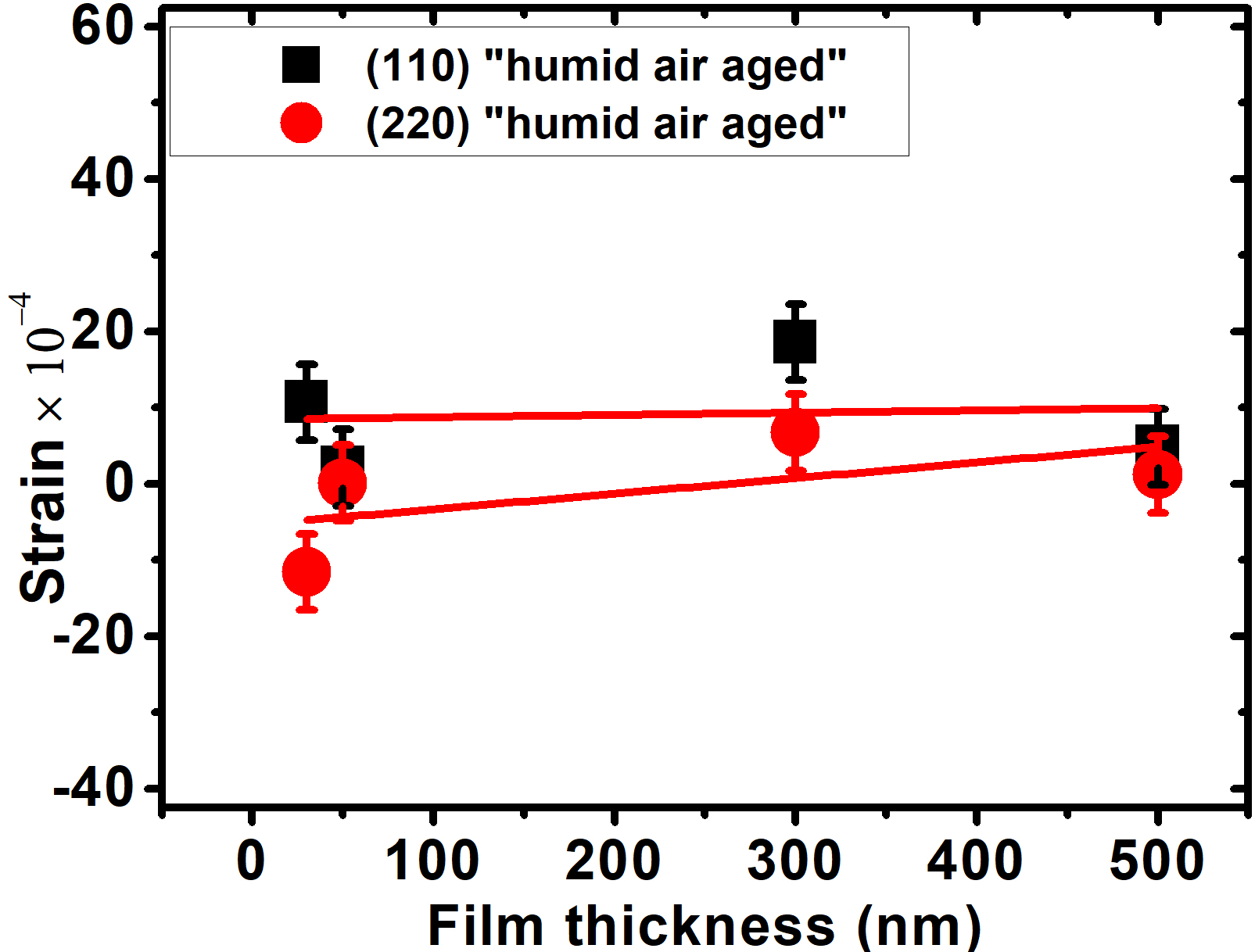}
	\caption{The variation of strain as a function of film thickness for (110) and (220) lattice planes in cases of "as-deposited" and "humid air aged" CsI thin films.}
\end{figure}

The lattice constant of CsI thin films has been calculated by using the analytical relation (for cubic crystal system)~\cite{B. D. Cullity}. The variation of the calculated lattice constants for all films is found in between $\sim$ 4.53 $\angstrom$ and $\sim$ 4.57 $\angstrom$ which are very close to the standard value (4.56 \angstrom) reported in JCPDS pdf no.: 060311 for single CsI crystal and to the experimentally observed values (4.55 \angstrom) from CsI powder.

Concerning the crystallinity of CsI thin films, the effect of film thickness on the texture coefficient ($TC_{(110)}$) of the most intense peak (110) and the crystallite size (D) on dislocation density ($\delta$) have also been studied.\\ 
The texture coefficient ($TC_{(110)}$) of the most intense peak (110) has been calculated by using equation (2)~\cite{11}:

	\begin{equation}
		\label{eq:emc}
		\mathbf{TC_{(110)} = \frac{I_{(110)}/I_{o(110)}}{N^{-1}\sum_N\frac{I_{(110)}}{I_{o(110)}}}}
	\end{equation}
Where $I_{(110)}$ is the measured intensity of (110) plane on the film, $I_{o(110)}$ is the intensity of (110) plane taken from JCPDS pdf no.: 060311 and N is the number of reflections taken into analysis. From Fig. 5a, one can observe that the texture coefficient of the most intense peak (110) ($TC_{(110)}$) for all CsI films has values more than unity ($TC_{(110)}\textgreater 1$). This indicates that the preferred growth of all CsI films is oriented along (110) crystallographic plane. Also it shows film thickness dependent, where the reflective films have higher values of $TC_{(110)}$ in comparison with the semitransparent films. The $TC_{(110)}$ increases from 1.32 to 3.09 as the film thickness increases from 30 nm to 500 nm. Moreover we found that the $TC_{(110)}$ is decreased after exposing to humidity (not shown in the figure). Hence, the increase in the intensity of (110) lattice plane as well as $TC_{(110)}$ simultaneously with the increase in film thickness (from semitransparent films to reflective films) provides the improvement of crystalline quality. 
	
	\begin{figure}[!ht]
		\centering
		\includegraphics[scale=0.28]{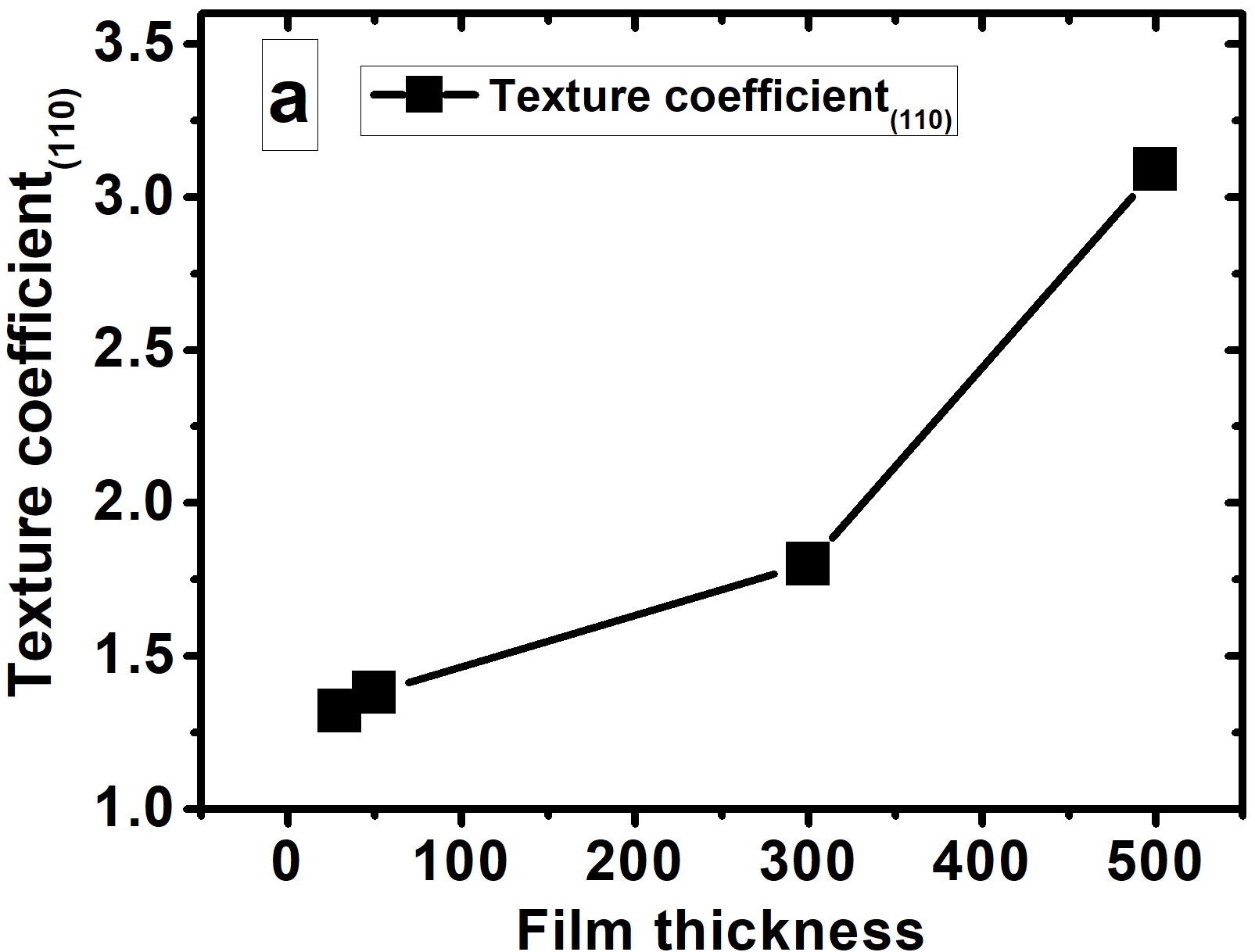}
		\hspace{0.5cm}
		\includegraphics[scale=0.28]{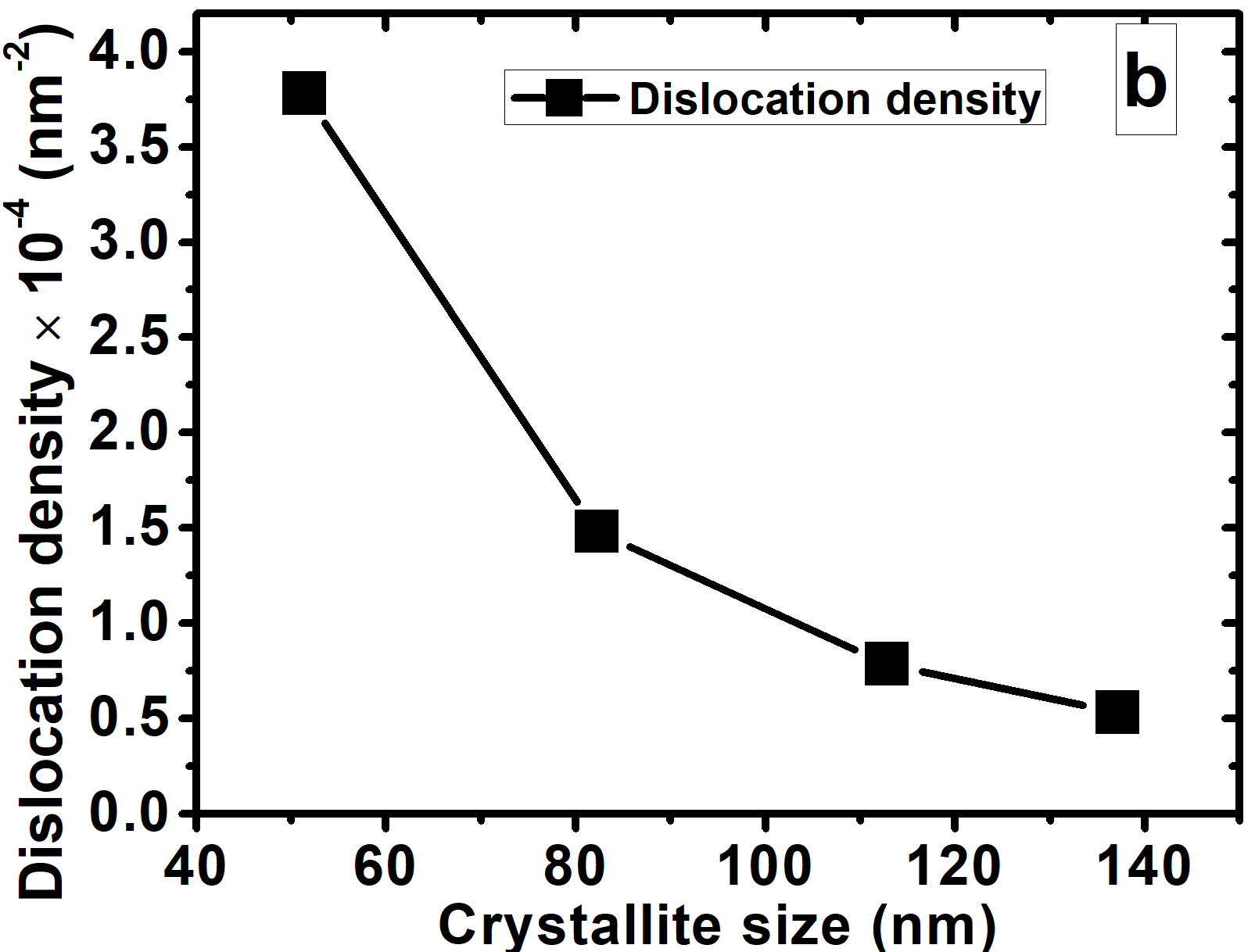}
		
		\caption{(a): The variation of texture coefficient as a function of film thickness and (b): The variation of dislocation density as a function of crystallite size for (110) lattice plane.}
	\end{figure} 
	
The dislocation density ($\delta$) defines as a measure of the number of dislocations in a unit volume of a crystalline material. It has been evaluated by using the values of crystallite size through the formula $\delta = 1/D^2$~\cite{12}. In Fig. 5b, it can be seen that the dislocation density decreases with the increase in crystallite size. While the crystallite size increases from 51 nm to 137 nm, the dislocation density decreases from $3.78\times10^{-4}~nm^{-2}$ to $1.48\times10^{-4}~nm^{-2}$. Further, we found that all CsI films show a decrease in dislocation density value after exposing to humidity. It is known that the reduction of dislocation density with the increase in crystallite size indicates enhancement of the film crystallinity~\cite{S. Benramache}. 
	
\subsection{\textbf{Optical properties of CsI thin films}}

The absorbance results of the reflective and semitransparent CsI thin films in case of "as-deposited" and "humid air aged" are displayed in Figs. 6a and 6b respectively. In both cases, the absorbance has maximum values in the ultraviolet region and start decreasing with the increase in wavelength to have minimum values in the visible and infrared region. The absorbance depends basically on the film thickness where it can be seen that the absorbance decreases with the decrease in film thickness. Moreover, one can observe that the absorbance decreases further after exposure to humidity as shown in Fig. 6b. At the wavelength $\lambda \sim 205$ nm, a nice peak has appeared with absorbance value about 4.2 for the thicker film (500 nm) and about 1.6, 0.6, 0.5 for 300 nm, 50 nm and 30 nm thicknesses of "as-deposited" CsI films respectively. The effect of humidity on the CsI films is manifested on the reflective films where the absorbance decreases to a maximum value of about 2.4 for the thicker film (500 nm), see Fig. 6b (inset). The optical absorption of reflective and semitransparent CsI thin films has been obtained by C. Lu et al.~\cite{Lu}, P. Maier et al.~\cite{P. Maier}, T. Boutboul et al.~\cite{T. Boutboul} and Triloki et al.~\cite{we}. We found that our results are having good match with them.
	
	\begin{figure}[!ht]
		\centering
		\includegraphics[scale=0.28]{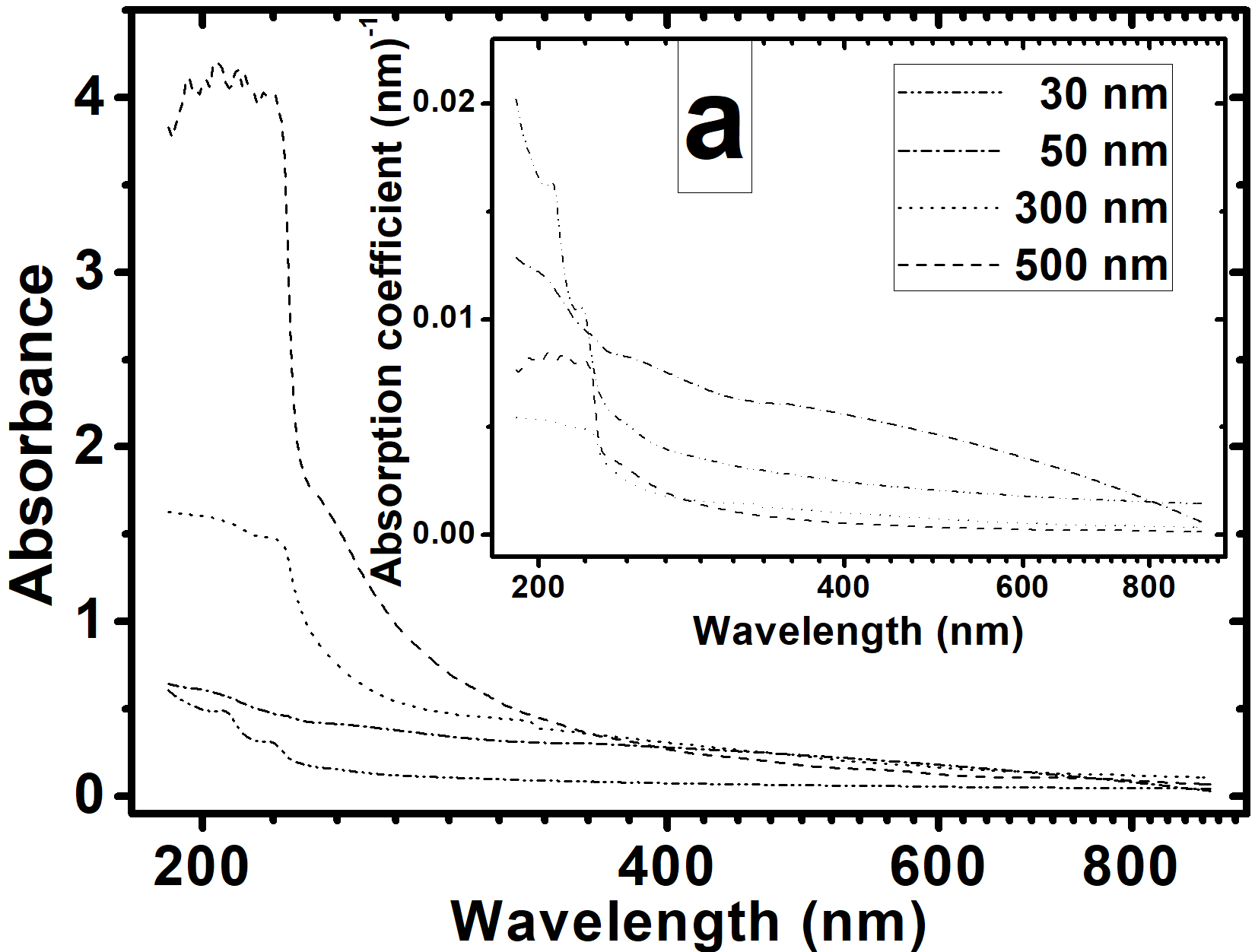}
		\hspace{0.5cm}
		\includegraphics[scale=0.28]{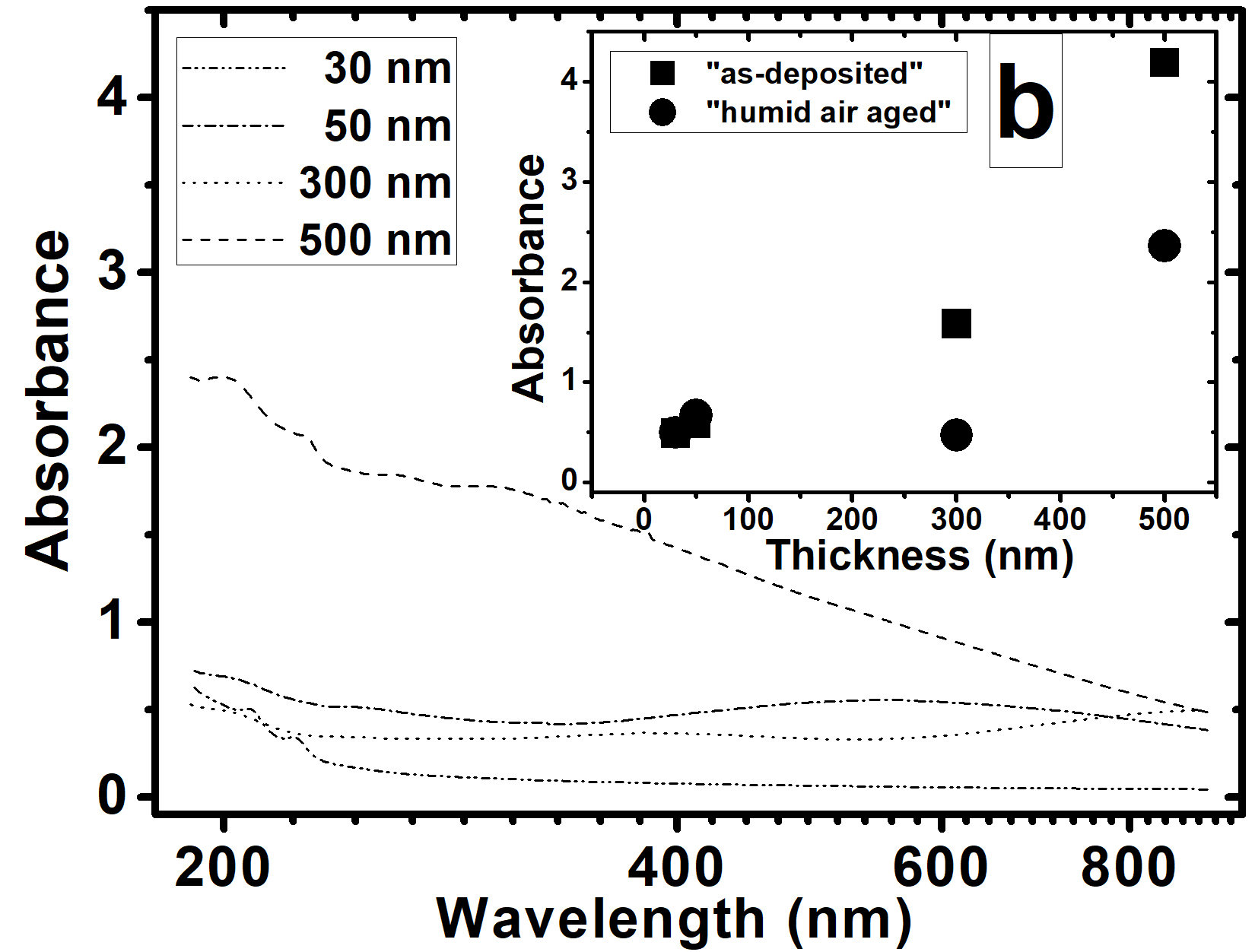}
		\caption{The optical absorbance of "as-deposited" and "humid air aged" CsI thin films as a function of wavelength. (a inset): The absorption coefficient of CsI thin films as a function of wavelength. (b inset): The variation of absorbance as a function of film thickness at the wavelength $\lambda \sim$ 205 nm.}
	\end{figure}

The relation connecting the absorption coefficient ($\alpha$), the incident photon energy ($h \nu$) and optical band gap (Eg) takes the form of Tauc relation of K.M. model {$(\alpha h \nu) ^ n = A (h\nu - Eg)$} where A is the edge width parameter, h is the Planck's constant, $\nu$ is the frequency of vibration and n is either equal to 2 for direct band transition or equal to 1/2 for indirect band transition~\cite{16,17}. The absorption coefficient ($\alpha$) has been calculated by the present absorption spectra and found to be in between 0.0001 $nm^{-1}$ and 0.0202 $nm^{-1}$, see Fig. 6a (inset). A rapid decrease in the absorption coefficient at the ultraviolet region near the absorption edge has also been found, which indicates the possibility of a direct transmission of electrons from valence to conduction band. The relation between $(\alpha h \nu)^ 2$ and $(h \nu)$ is used to determine the band gap of the film and by drawing a linear interpolation of each curve to energy axis, the band gap energy (Eg) can be estimated as shown in Fig. 7a. We have found that the band gap energies of 500 nm, 300 nm, 50 nm and 30 nm CsI films are about 5.30 eV, 5.25 eV, 5.00 eV and 5.20 eV respectively. A. Buzulutskov et al.~\cite{18} have also derived the band gap energy to be 5.90 eV from experimental QE dependence on wavelength. The high band gap energy obtained by A. Buzulutskov et al.~\cite{18} may be due to the heat treatment of CsI film compared to "as-deposited" films in our case. Recently, Triloki et al.~\cite{Trilokietal} have also obtained the band gap energy of 500 nm thick CsI film by Tauc relation to be about 5.40 eV. In the present work, the estimated band gap energy of 500 nm thickness CsI thin films is about 5.30 eV. This value is in close agreement with Triloki et al.~\cite{Trilokietal}. The estimated band gap energies of reflective and semitransparent CsI films show thickness dependence as shown in Fig. 7b. It is easy to observe that the band gap energy is slightly increased from semitransparent to reflective films. The film thickness is proportional to the size of the particle in the crystallites of thin film. As the crystallite size affects the energy level of electrons, the band gap will be dependent on film thickness. Further, this variation of band gap energy with film thickness is caused by the variation of the film disorder, where it increases in the thinner films~\cite{Sonmezoglu,Ovshinsky}.
	
	\begin{figure}[!ht]
		\centering
		\includegraphics[scale=0.28]{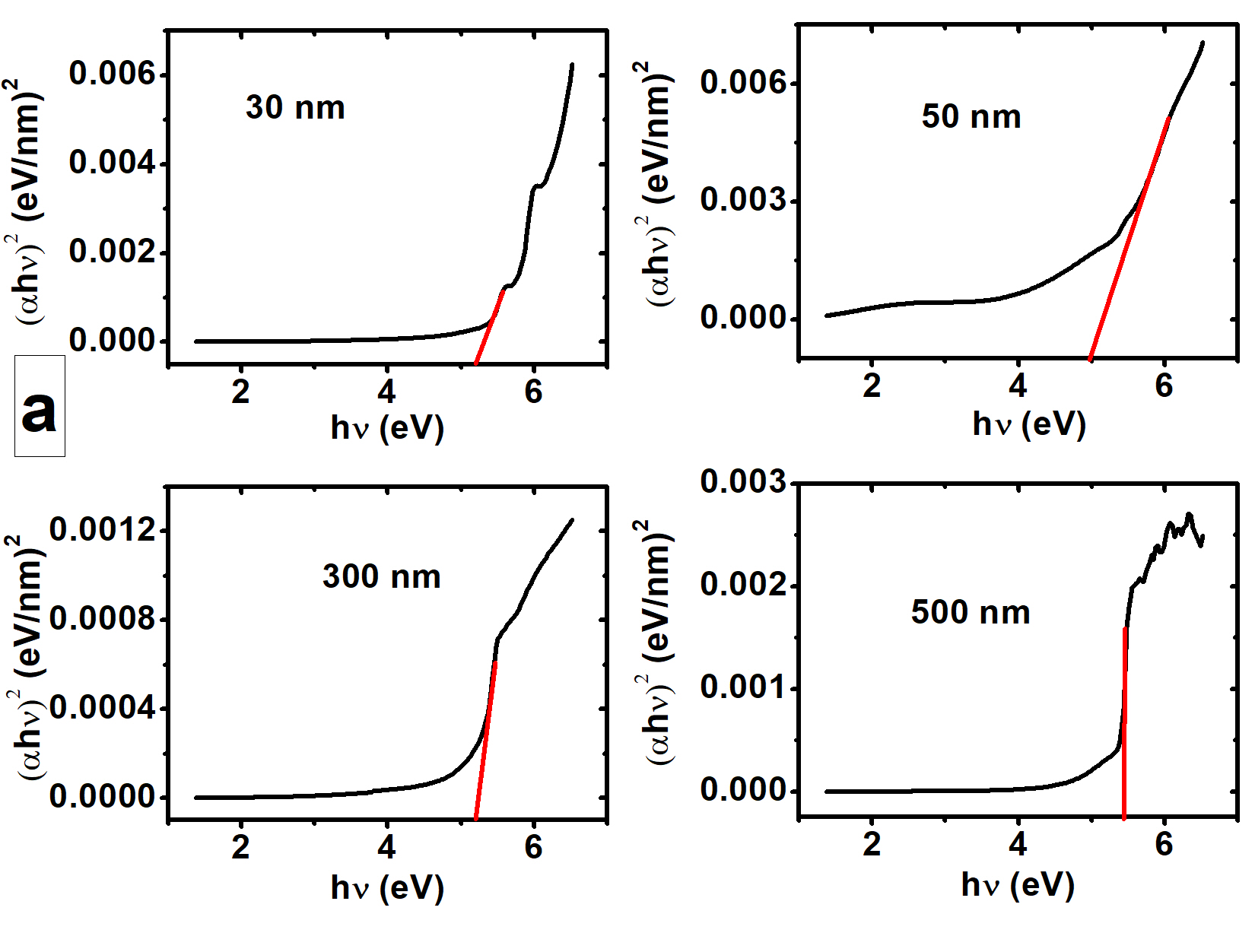}
		\hspace{0.5cm}
		\includegraphics[scale=0.28]{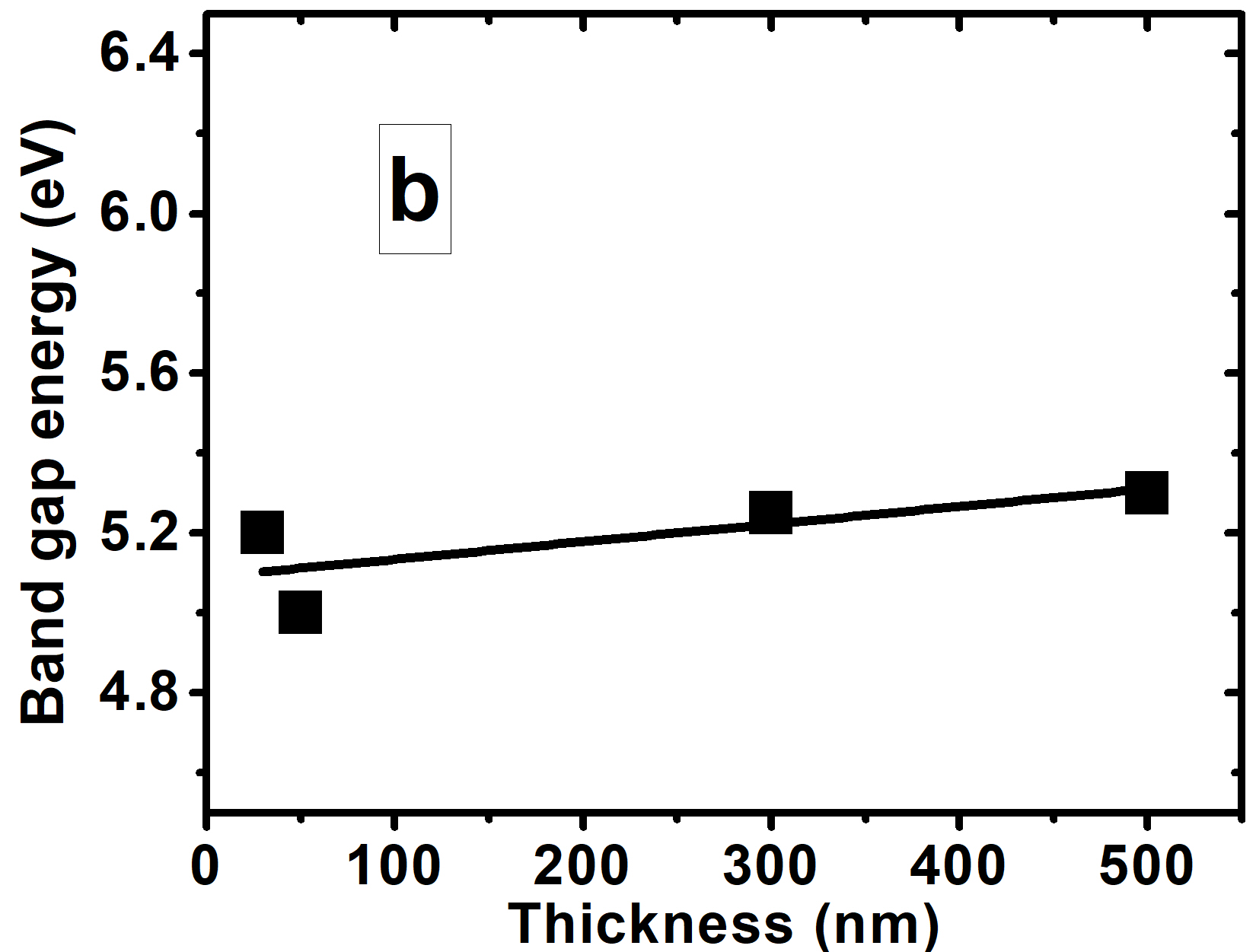}
		\caption{(a): Variation of $(\alpha h \nu)^ 2$ vs. photon energy $(h \nu)$ for all CsI films and (b): The band gap energy as a function of film thickness.}
	\end{figure}
	
The reasons of variation the band gap energy with film thickness in the thin films could be one or more of the following effects: (1) The change in barrier height at grain boundaries with film thickness. (2) High density of dislocation and (3) Quantum size effect. The dislocation density effects in the band gap of material when it is very high that causes expansion in the spacing of the atoms. In this investigation, one may note that the values of dislocation density are in the factor of ($10^{-4}$) as shown in Fig. 5b which can be considered as not too high, also it can be confirmed by the narrow peaks of XRD diffraction. Thus, the contribution of dislocation density in the variation of band gap may be small. The quantum size effect can be dismissed for reflective films (500 nm and 300 nm) however for semitransparent films (50 nm and 30 nm), it may not be ignored. Therefore, the first effect has motivated us for further investigation. In thin films, if the charges are accumulated at grain boundaries, the band gap energy may increase. 
	
According to Slater's~\cite{23} proposal; the energy barrier, grain boundaries and barrier height can be affected by the charge accumulation at grain boundaries, so the correlation between the grain boundary barrier height and crystallite size can be known by using equation (3):
	\begin{equation}
		\label{eq:emc}
		\mathbf{E_b = E_{bo} + C(X-fd)^2}
	\end{equation}
Where E$_b$ is the energy barrier, E$_{bo}$ is the original barrier height, C is a constant depending on the density of charge carriers and dielectric constant of the material, X is the barrier width (20 nm - 30 nm), d is the crystallite size and f is a fraction of the order of 1/15 to 1/50 depending on the charge accumulation and carrier concentration. To consider the low carrier concentration of the most intense peak (110) of CsI thin films, we take X= 26 nm and f= 1/23. The comparison between the variation of the calculated factor (X - fd)$^2$ as a function of crystallite size with the variation of energy gap with crystallite size are shown in Fig. 8. The good agreement indicates the influence of barrier height contribution to the observed optical energy gap variation with film thickness. Moreover, it can be seen that there is an inverse dependence between the estimated optical band gap energy (obtained from K. M. model) and crystallite size (obtained from (110) XRD lattice plane). This dependence is typically found because of quantum confinement effect associated to the films crystallites in very small nanoparticles~\cite{E.M. Wong,L. Banyai}.

	\begin{figure}[!ht]
		\centering
		\includegraphics[scale=0.3]{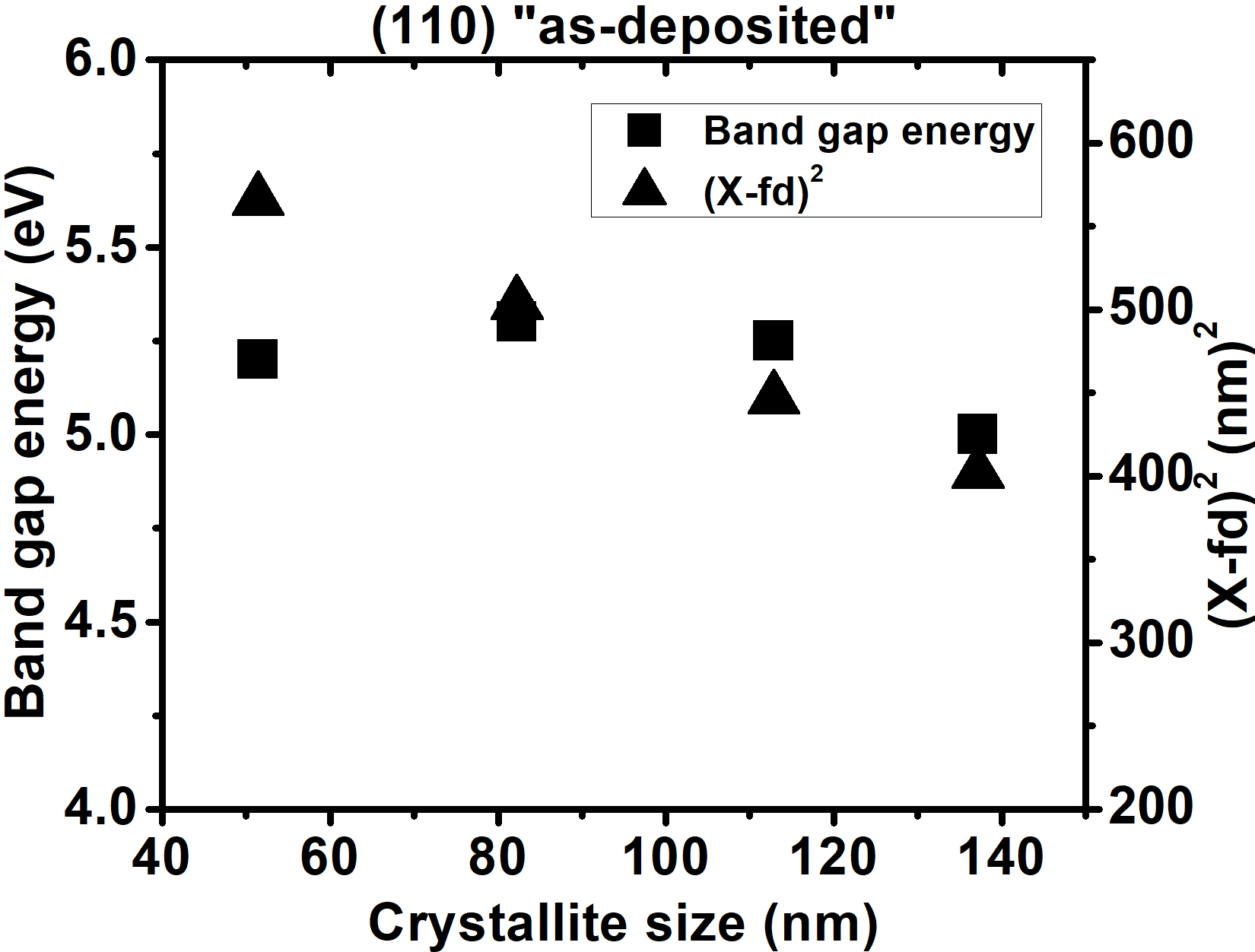}
		\caption{The dependence of optical energy gap along with the estimated grain boundary barrier height variation factor $(X-fd)^2$ as a function of crystallite size.}
	\end{figure}

Fig. 9 shows the relation between the band gap energy and lattice constant for (110) and (220) lattice planes. It can be observed that the band gap energy for both lattice planes decreases with the increase in lattice constant. The reason behind that could be the movement between the Cs and I atoms away from each other which results increasing in lattice constant and this might be another reason for the variation in the estimated band gap energy.
	
	\begin{figure}[!ht]
		\centering
		\includegraphics[scale=0.3]{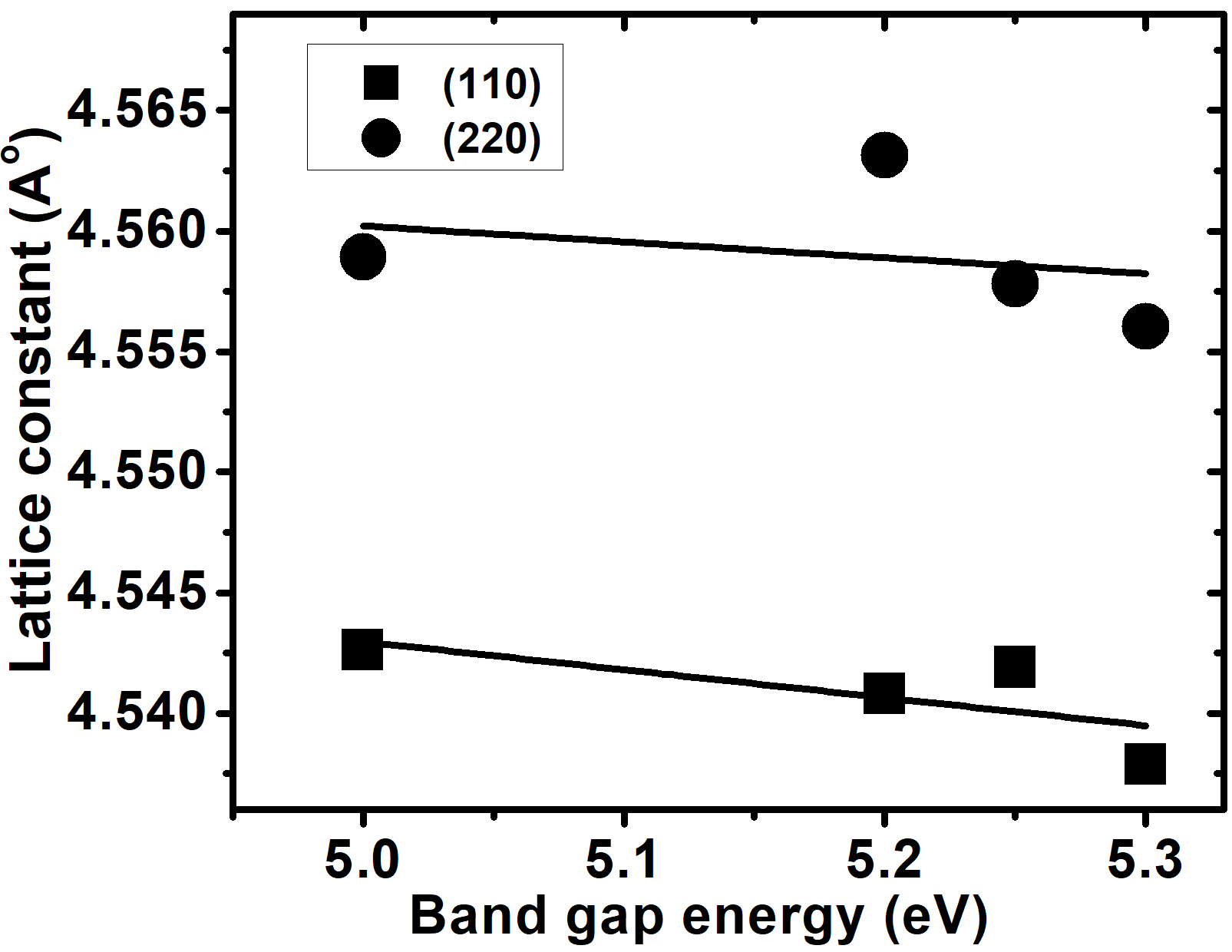}
		\caption{The lattice constant (a) as a function of band gap energy (Eg) for (110) and (220) lattice planes.}
	\end{figure}

\section{Conclusion}
In conclusion, CsI thin films with reflective and semitransparent thicknesses are deposited on stainless steel and quartz substrates by using the thermal evaporation technique. The lattice plane assigned to the most intense peak for all CsI films in case of "as-deposited" and "humid air aged" films is mainly (110) and the lattice plane (220) is found in mutual for all CsI films. The peaks are shifted to the higher angles of $2\theta$ compared to its position in case of CsI powder. This shift indicates a compressive stress developed in the films. The compressive strain is found to increase slightly with the increase in film thickness. There is no effect of humidity on the compressive strain values where it is found to be in the same range of the "as-deposited" one, therefore, it may arise due to some other factors. The lattice constant is found to vary from $\sim$~4.53 $\angstrom$ to $\sim$~4.75 $\angstrom$ which is very close to the standard value 4.56 $\angstrom$ and to the experimentally observed values (4.55 $\angstrom$) from CsI powder. The crystallinity of CsI films is found to be improved with the increase in sharpness of XRD peaks as well as with the decrease in dislocation density simultaneously with the increase in film thickness.
	
The optical absorbance is found to be film thickness dependence. The effect of humidity on optical absorbance is found to be evident in the reflective film. The absorbance of 500 nm thick CsI film is decreased after exposing to humidity from 4.2 to 2.4 at the wavelength $\sim$ 205 nm, and the film becomes more transparent. K. M. model has been used to estimate the optical band gap energy of CsI films. The band gap energies of 500 nm, 300 nm, 50 nm and 30 nm CsI films are found to be about 5.30 eV, 5.25 eV, 5.00 eV and 5.20 eV respectively. The observed discrepancy with and Buzulutskovet al.~\cite{18} data (Eg$\sim$ 5.9 eV) may be due to the heat-enhanced CsI film compared to "as-deposited" CsI film in our study. It is found that the values of band gap energy vary with the film thickness. This variation is attributed to the variation of film disorder and the change in barrier height at grain boundaries with film thickness. According to Slater's proposal, the barrier height contribution is found to influence the observed optical energy gap variation with film thickness.

\section*{Acknowledgment}
	
This work was partially supported by the FIST, PURSE of Department of Science and Technology (DST), CAS program of University Grant Commission (UGC) and by the Indian Space Research Organization (ISRO) SSPS programs, Government of India.

\end{document}